\documentclass[a4paper]{jpconf}
\usepackage{amsmath}
\usepackage{graphicx}
\usepackage{hyperref}
\usepackage{bm}\renewcommand{\mathbf}{\bm}

\newcommand{\figwidth}{0.95\columnwidth}

\begin{document}

\title{Equations of Motion for the Out-of-Equilibrium Dynamics of Isolated Quantum Systems from the Projection Operator Technique}

\author{N. Nessi$^1$ and A. Iucci$^1$ }
\address{
$^1$Instituto de F\'{\i}sica La Plata (IFLP) - CONICET and Departamento de F\'{\i}sica,\\
Universidad Nacional de La Plata, CC 67, 1900 La Plata, Argentina}

\ead{nnessi@fisica.unlp.edu.ar}

\begin{abstract}
We present a rigorous framework to obtain evolution equations for the momentum distribution and higher order correlation functions in weakly interacting systems based on the Projection Operator Technique. These equations can be numerically solved in an efficient way. We compare the solution of the equations with known results for 1D models and find an excellent agreement.\end{abstract}

\section{Introduction}

The out-of-equilibrium dynamics of closed quantum systems is nowadays a very active field of research (see~\cite{polkovnikov11_nonequilibrium_dynamics} for a review). The central question behind these efforts is to describe the dynamics of an isolated many-body system towards an eventual stationary state after it has been driven out of equilibrium. Of particular interest are the properties of the equilibrium state itself. If simple correlation functions in this equilibrium state can be reproduced by means of a canonical ensemble the system is said to thermalize, which, interestingly, has been shown not to occur in certain situations~\cite{kinoshita06_non_thermalization,rigol07_generalized_gibbs_hcbosons}. These issues, that lie at the heart of statistical mechanics, have recently received a very strong impulse from the experiments with cold atomic gases. These systems are almost perfectly isolated from the environment and can remain quantum coherent for a long time compared with typical experiments duration allowing to study \textit{in situ} the dynamics of large aggregates of interacting quantum particles~\cite{kinoshita06_non_thermalization}.

From the theoretical perspective, the non-equilibrium dynamics of large, interacting many-body systems settles a formidable challenge. Among the several theoretical approaches at disposal, different kinds of evolution equations for few-body correlations are being widely employed in the contemporary literature (see, for example~\cite{tavora13_kinetic_bosons_quench,furst13_hubbard_matrix_boltzmann,stark13_kinetic_description,berges04_nonequilibrium_field_theory}).  In contrast to numerical methods, which are confined either to small systems or short times, as for instance exact diagonalization or time-dependent density matrix renormalization group (t-DMRG), the solution of evolution equations allows to investigate the dynamics of large systems for relatively long times and usually demands a rather modest numerical effort. However, the deduction of these kind of equations starting from first principles, i.e., the many-body Schr\"{o}dinger equation or its equivalents, often involves a series of uncontrolled approximations~\cite{berges04_nonequilibrium_field_theory,berges06_validity_transp_eqs} that, although routinely utilized, cannot be justified in all cases.

On the other hand, the most studied type of evolution equations are kinetic equations, which are markovian, i.e. memoryless, equations for one-particle distribution functions, a typical example being the Boltzmann equation. Even though the markovian limit is well defined in some situations~\cite{grabert82_book_pot,rau95_review_pot} it may not be always justified. In particular, in situations in which the system is reluctant to loose memory of the initial conditions (such as integrable models) markovian kinetic equations may be loosing important features of the relaxation dynamics.

In this work we present a method for constructing evolution equations for few body correlation functions in weakly interacting systems that involves no approximations beyond a series expansion in the small parameter quantifying the strength of the interaction. The resulting equations of motion are inherently non-markovian. The approach is based on the Projection Operator Technique (POT)~\cite{grabert82_book_pot,rau95_review_pot,breuer82_book_open_systems}. We benchmark the method by comparing the solution of the equations with known results from one dimensional (1D) systems.

\section{Model and statement of the problem}

We consider a system of interacting spinless fermions with Hamiltonian $H=H_0+\alpha H_1$ with
\begin{equation}
H=H_0+\alpha H_1=\sum_{\mathbf{k}}\epsilon(\mathbf{k})n(\mathbf{k})+\alpha\sum_{\mathbf{k}_{1},\mathbf{k}_{2},\mathbf{k}_{3},\mathbf{k}_{4}}V^{\mathbf{k}_{1},\mathbf{k}_{2}}_{\mathbf{k}_{3},\mathbf{k}_{4}}c^{\dagger}(\mathbf{k}_{1})c^{\dagger}(\mathbf{k}_{2})c(\mathbf{k}_{3})c(\mathbf{k}_{4}),
\end{equation}
where $c^{\dagger}(\mathbf{k})$ and $c(\mathbf{k})$ are fermionic creation and annihilation operators satisfying canonical anticommutation relations, $V^{\mathbf{k}_{1},\mathbf{k}_{2}}_{\mathbf{k}_{3},\mathbf{k}_{4}}$ is the momentum-space matrix element of the interaction, $\epsilon(\mathbf{k})$ is the dispersion relation, $n(\mathbf{k})=c^{\dagger}(\mathbf{k})c(\mathbf{k})$ is the number operator and $\alpha$ is the strength of the interaction. Our results can be easily extended to the bosonic case. The hermiticity of the Hamiltonian and the symmetry in the sum indices impose $V^{\mathbf{k}_{1},\mathbf{k}_{2}}_{\mathbf{k}_{3},\mathbf{k}_{4}}=-V^{\mathbf{k}_{2},\mathbf{k}_{1}}_{\mathbf{k}_{3},\mathbf{k}_{4}}=-V^{\mathbf{k}_{1},\mathbf{k}_{2}}_{\mathbf{k}_{4},\mathbf{k}_{3}}=\bar{V}^{\mathbf{k}_{4},\mathbf{k}_{3}}_{\mathbf{k}_{2},\mathbf{k}_{1}}$, where $\bar{V}$ denotes the complex conjugate.

Furthermore, we will be interested in the special case of a translationally invariant Hamiltonian in which the particles interact via a pair potential $v(\mathbf{x}-\mathbf{y})$. In such case
\begin{equation}
V^{\mathbf{k}_{1},\mathbf{k}_{2}}_{\mathbf{k}_{3},\mathbf{k}_{4}}=\frac{1}{4V}\delta_{\mathbf{k}_{1}+\mathbf{k}_{2},\mathbf{k}_{3}+\mathbf{k}_{4}}\left(\hat{v}(\mathbf{k}_{1}-\mathbf{k}_{4})-\hat{v}(\mathbf{k}_{2}-\mathbf{k}_{4})-\hat{v}(\mathbf{k}_{1}-\mathbf{k}_{3})+\hat{v}(\mathbf{k}_{2}-\mathbf{k}_{3})\right),
\end{equation}
 where $\hat{v}(\mathbf{k})=\int d\mathbf{r}\, v(\mathbf{r})e^{i\mathbf{k}\cdot\mathbf{r}}$ is the Fourier transform of the potential and we have written the antisymmetrized version in order to respect the symmetry conditions of the potential. We are interested in the evolution of the system starting from an arbitrary initial condition given by a density matrix $\rho(0)$ which we leave unspecified until the next section.

\section{Evolution equation for the momentum distribution}

Our starting point is the Liouville equation in the interaction representation ($\hbar$=1):
\begin{equation}\label{eq:liouville}
\partial_{t}\tilde{\rho}(t)=-i\alpha[\tilde{H}_{1}(t),\tilde{\rho}(t)]=\alpha L(t)\tilde{\rho}(t),
\end{equation}
where $\tilde{O}(t)=e^{iH_{0}t}Oe^{-iH_{0}t}$ is the interaction representation of the operator $O$ and we have introduced the Liouville superoperator $L(t)O=-i[\tilde{H}_{1}(t),O]$. Our task is to find approximate solutions to the microscopic dynamics described by the Liouville equation. The POT defines a program for achieving this. We need to first identify the ``slow'' or ``macroscopic'' variables in our system and then project the dynamics into the subspace of these slow variables. In a weakly interacting system the occupation number operators emerge as natural slow variables since $[H,n(\mathbf{k})]=\mathcal{O}(\alpha)$. To perform the projection we first introduce the ``relevant'' density matrix
\begin{equation}
\sigma(t)=\frac{1}{Z(t)}\exp\left[-\sum_{\mathbf{k}}\lambda(\mathbf{k},t)n(\mathbf{k})\right],
\end{equation}
where the time-dependent partition function is given by $Z(t)=\mathrm{Tr}\left[\exp\left(-\sum_{\mathbf{k}}\lambda(\mathbf{k},t)n(\mathbf{k})\right)\right]$. Note that $\tilde{\sigma}(t)=\sigma(t)$. The Lagrange multipliers $\lambda(\mathbf{k},t)$ enforce the relation:
\begin{equation}\label{eq:constraints}
\langle n(\mathbf{k})\rangle_{t}\equiv \mathrm{Tr}[n(\mathbf{k})\sigma(t)]=\mathrm{Tr}[n(\mathbf{k})\rho(t)].
\end{equation}
In other words, $\sigma(t)$ is the density matrix that maximizes the Von Neumann entropy subject to the constraints~\eqref{eq:constraints}. We note that although the dynamics of $\tilde{\rho}(t)$ is generated by the Liouville equation, the relevant density matrix $\sigma(t)$ is time-dependent independently of the representation. The projection of the dynamics consists in finding an equation of motion for $\sigma(t)$. To this end we introduce a projection super-operator $P(t)$ that projects the relevant density matrix $P(t)\tilde{\rho}(t)=\tilde{\sigma}(t)$. We refer the interested reader to the literature to find an explicit expression for $P(t)$~\cite{grabert82_book_pot,rau95_review_pot,breuer82_book_open_systems,robertson66_pot_first,kawasaki73_non-linear_transport}.
%\begin{equation}\label{eq:projector}
%P(t)\mu=\left(\sigma(t)-\sum_{\mathbf{k}}\frac{\delta\sigma(t)}{\delta\langle n(\mathbf{k})\rangle_{t}}\langle n(\mathbf{k})\rangle_{t}\right)\mathrm{Tr}\left[\mu\right]+\sum_{\mathbf{k}}\frac{\delta\sigma(t)}{\delta\langle n(\mathbf{k})\rangle_{t}}\mathrm{Tr}\left[n(\mathbf{k})\mu\right],
%\end{equation}
%where $\mu$ is an arbitrary operator. The projection operator Eq.~\eqref{eq:projector} projects the relevant density matrix $P(t)\tilde{\rho}(t)=\tilde{\sigma}(t)$ and, besides the usual properties~\cite{grabert82_book_pot,rau95_review_pot,breuer82_book_open_systems}, due to momentum conservation it also satisfies $P(t)L(t)P(s)=0$.
%\begin{eqnarray}\label{eq:p_properties}
%\nonumber P(t)\tilde{\rho}(t)&=&\tilde{\sigma}(t),\\
%\nonumber P(t)\partial_{t}\tilde{\rho}(t)&=&\partial_{t}\tilde{\sigma}(t),\\
%\nonumber \mathrm{Tr}\left[n(\mathbf{k})P(t)\mu\right]&=&\mathrm{Tr}\left[n(\mathbf{k})\mu\right],
%\nonumber \\P(t)P(t')\mu&=&P(t)\mu,\\
%P(t)L(t)P(s)\mu&=&0.
%\end{eqnarray}
 %The fourth identity, setting $t=t'$, expresses the idempotent character of the projector whereas the last identity depends on the explicit form of the Hamiltonian $H$.
 It is also useful to define the complementary projector $Q(t)=1-P(t)$.

 Following the usual steps~\cite{grabert82_book_pot,rau95_review_pot,breuer82_book_open_systems,robertson66_pot_first}, from the Liouville equation we obtain an equation for the slow dynamics %$\partial_{t}P(t)\tilde{\rho}(t)=\alpha P(t)L(t)\tilde{\rho}(t)$ and other for the dynamics of the fast, ``microscopic'' degrees of freedom $\partial_{t}Q(t)\tilde{\rho}(t)=\alpha Q(t)L(t)\tilde{\rho}(t)$ and then formally solve the equation of the fast degrees of freedom and insert the result into the slow part to obtain
 \begin{equation}\label{eq:robertson}
 \partial_{t}\tilde{\sigma}(t)=\alpha P(t)L(t)\tilde{\sigma}(t)+\alpha^{2}\int_{0}^{t}ds\, P(t)L(t)G(t,s)Q(s)L(s)\tilde{\sigma}(s)+\alpha P(t)L(t)G(t,0)Q(0)\tilde{\rho}(0),
 \end{equation}
 where $G(s,t)=\mathrm{T_{\rightarrow}}\exp\left[-\alpha\int_{s}^{t}ds'\, Q(s')L(s')\right]$ is an anti-chronologically ordered exponential. The first term in Eq.~\eqref{eq:robertson} is a mean-field term that vanishes due to momentum conservation, the second one is a memory term that can be completely expressed in terms of the past history of the $\langle n(\mathbf{k})\rangle_{t}$'s and the third term is a microscopic noise term that cannot be expressed solely as a function of the slow variables. This last term vanishes if we choose an initial condition that has the same form as the relevant density matrices, i.e., if $\rho(0)=\sigma(0)$, and we shall in the following circumscribe to this case.

 Eq.~\eqref{eq:robertson} is equivalent to the Liouville dynamics and, in general, as difficult to solve as the original problem. It sets, however, a good starting point for approximations. To render Eq.~\eqref{eq:robertson} tractable we perform a perturbative expansion in the interaction strength using that $G(t,s)=I+\mathcal{O}(\alpha)$. Taking the trace $\langle n(\mathbf{k})\rangle_{t}=\mathrm{Tr}[n(\mathbf{k})\sigma(t)]$ we finally obtain
 \begin{equation}\label{eq:pert_rob}
 \partial_{t}\langle n(\mathbf{k})\rangle_{t}=\alpha^{2}\int_{0}^{t}ds\,\mathrm{Tr}\left[n(\mathbf{k})L(t)L(s)\tilde{\sigma}(s)\right]+\mathcal{O}(\alpha^{3}).
 \end{equation}
A great simplification arises since, given the Gaussian structure of $\sigma(t)$, we can use the Wick pairing rule to evaluate the trace in~\eqref{eq:pert_rob}. After a straightforward (but potentially tedious) calculation we obtain the explicit equation of motion
\begin{eqnarray}\label{eq:kinetic}
\nonumber f(\mathbf{k},t)&=&f(\mathbf{k},0)-16\alpha^{2}\sum_{\mathbf{k}_{2},\mathbf{k}_{3},\mathbf{k}_{4}}\vert V^{\mathbf{k},\mathbf{k}_{2}}_{\mathbf{k}_{3},\mathbf{k}_{4}}\vert^{2}\int_{0}^{t}ds\,\frac{\sin\left[(t-s)\Delta e^{\mathbf{k},\mathbf{k}_2}_{\mathbf{k}_3,\mathbf{k}_4}\right]}{\Delta e^{\mathbf{k},\mathbf{k}_2}_{\mathbf{k}_3,\mathbf{k}_4}}\\
&\times&\left(f(\mathbf{k},s)f(\mathbf{k}_{2},s)\bar{f}(\mathbf{k}_{3},s)\bar{f}(\mathbf{k}_{4},s)-f(\mathbf{k}_{3},s)f(\mathbf{k}_{4},s)\bar{f}(\mathbf{k},s)\bar{f}(\mathbf{k}_{2},s)\right)+\mathcal{O}(\alpha^{3}),
\end{eqnarray}
where $\Delta e^{\mathbf{k},\mathbf{k}_2}_{\mathbf{k}_3,\mathbf{k}_4}=\epsilon(\mathbf{k})+\epsilon(\mathbf{k}_{2})-\epsilon(\mathbf{k}_{3})-\epsilon(\mathbf{k}_{4})$ and, in order to ease the notation, we have defined $f(\mathbf{k},t)\equiv\langle n(\mathbf{k})\rangle_{t}$ and $\bar{f}(\mathbf{k},t)\equiv 1-\langle n(\mathbf{k})\rangle_{t}$. This equation, in slightly different versions, has appeared many times in the literature. In Ref.~\cite{rau95_review_pot} it was derived using the same tools that we present here but it was used only as an intermediate step to derive the Boltzmann equation whereas in Refs.~\cite{stark13_kinetic_description,erdos04_qbe} it was heuristically derived and used to study the dynamics of infinite dimensional models and to derive a quantum version of the Boltzmann equation, respectively. We want to stress that~\eqref{eq:kinetic} is valid also for lattice systems in which only quasi-momentum is conserved.

Although the approach is clearly perturbative, the results go beyond conventional lowest order perturbation theory because the
perturbation expansion is performed inside the integro-differential equations and, therefore, the coupling is involved in a highly non-linear way in the final
expressions. To asses the accuracy of the approximation is thus not a straightforward task. One possible alternative is to calculate higher order corrections to Eq.~\eqref{eq:kinetic}, which can be done systematically within the context of the projection operator technique. Then, the error could be calculated as the relative difference between the lowest order and the next-to-leading order solutions~\cite{breuer82_book_open_systems}. We relegate such error analysis to future work. An alternative approach is to compare results with other techniques. For instance, performing a suitable short-times approximation on~\eqref{eq:kinetic} (details can be found in~\cite{stark13_kinetic_description}) it is possible to rederive results first obtained by Moeckel and Kehrein using the flow-equation approach in lowest order perturbation theory~\cite{moeckel08_quench_hubbard_high_d}. Moreover, in~\cite{stark13_kinetic_description} it was found that the solution of Eq.~\eqref{eq:kinetic} compares well on the accessible timescales with dynamical mean-field theory results for the infinite dimensional Hubbard model. Below we provide an example for a 1D model in which the solution of the equation~\eqref{eq:kinetic} captures highly non trivial features of the short-to-intermediate times relaxation. Regarding the validity of the approach for longer timescales than those accessible for current numerical techniques, we should mention that under certain assumptions it can be shown~\cite{erdos04_qbe} that Eq.~\eqref{eq:kinetic} reduces, for long enough times, to the quantum Boltzmann equation, which is expected to successfully describe the thermalization of isolated weakly interacting quantum systems after the system has effectively lost the information about the details of the initial conditions~\cite{berges06_validity_transp_eqs}.

With respect to implementation details, Eq.~\eqref{eq:kinetic} can be solved using standard techniques for systems of Volterra integral equations~\cite{linz85_volterra_eqs}. A straightforward algorithm for the solution using, for instance, the trapezoidal rule to perform the time integral, implies a calculation time that scales as $L^{3D}\times N^2$, where $N$ is the number of times steps and $D$ the space dimension. We have found an algorithm whose execution time scales as $L^{3D}\times N$ allowing us to reach large sizes and times. We finally note that the evolution equations are very suitable for parallel computing.

\subsection{Results for a 1D model}

\begin{figure}[t]
  \includegraphics[width=\figwidth]{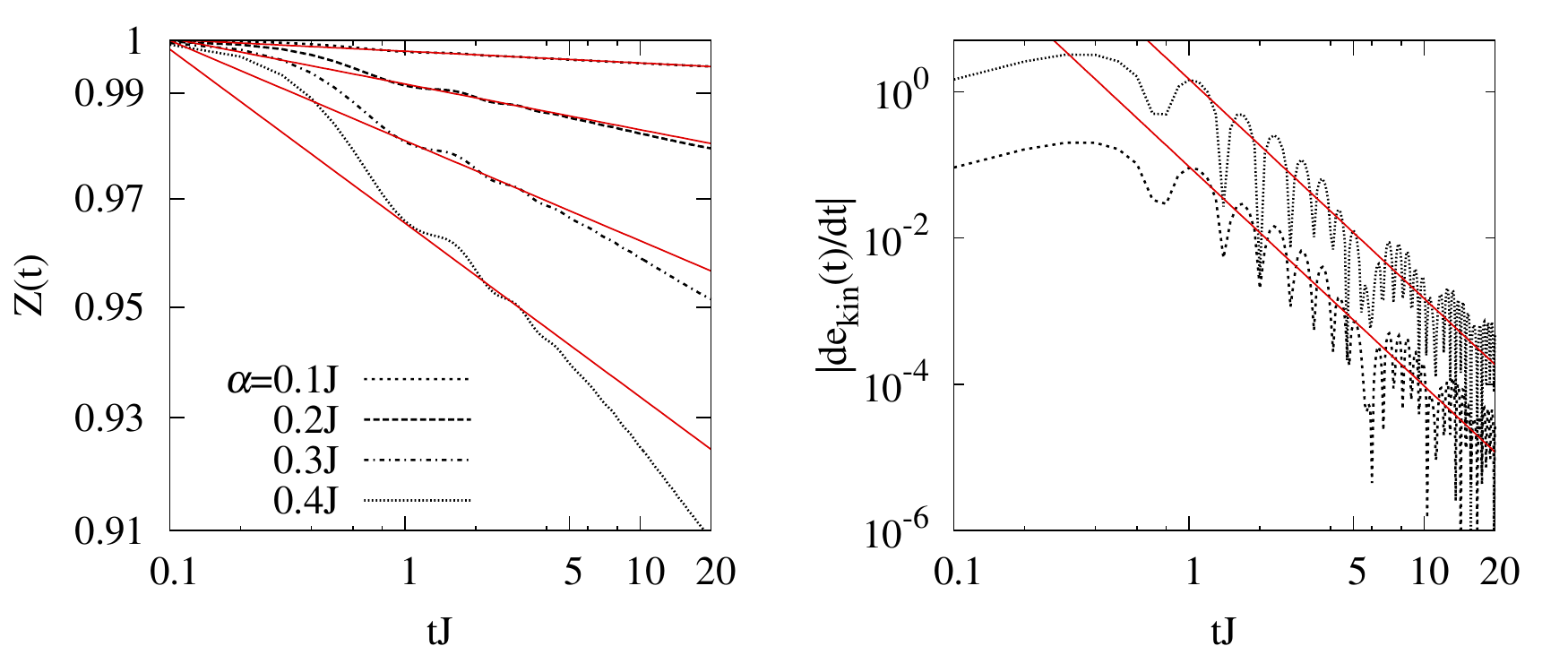}\\
  \caption{Right panel: Discontinuity in momentum distribution at the Fermi momentum,
   $Z(t)$. Solid lines are power laws with the exponent given by the LM predictions $Z(t)\sim t^{-\gamma}$, with $\gamma=\frac{1}{4}(K^{2}+K^{-2}-2)$. The Luttinger parameter is obtained from bosonization, $K=\sqrt{\frac{\pi v_{F}-\alpha}{\pi v_{F}+3\alpha}}$. Left panel: Derivative of the kinetic energy $e_{kin}(t)$. Solid lines are power laws $t^{-3}$ given by the LM predictions. The system size is $L=256$.}\label{fig:graph}
\end{figure}

Now we present results obtained using~\eqref{eq:kinetic} in a 1D model of spinless fermions with nearest-neighbor hopping and interactions
\begin{equation}
H=-J\sum_{j}(c_{j}^{\dagger}c_{j+1}+c_{j+1}^{\dagger}c_{j})+\alpha\sum_{j}n_{j}n_{j+1},
\end{equation}
where $n_j=c^{\dagger}_jc_j$ and sums run for $j=0,L-1$ ($L$ is the length of the system) and we impose periodic boundary conditions. This model can be mapped on to the XXZ model by means of a Jordan-Wigner transformation and, in equilibrium, exhibits a Luttinger liquid phase for $\alpha<2J$.  We study a system at half-filling ($k_F=\frac{\pi}{2}$) and take as initial state the ground state of the Hamiltonian with $\alpha=0$, i.e., we consider the ``quantum quench'' scenario. The same situation was studied in Ref.~\cite{karrasch12_ll_universality_quench} using t-DMRG where, remarkably, it was found that the short time dynamics were accurately described by the quench dynamics of the Luttinger model (LM)~\cite{cazalilla06_quench_LL,iucci09_quench_LL}, raising the question of to what extent the equilibrium low energy fixed point of a given model can describe the out of equilibrium dynamics. In Fig.~\ref{fig:graph} we show results for two different observables, the jump $Z(t)$ of the momentum distribution at $k_F$ and the kinetic energy $e_{kin}(t)$,  and compare them with the LM predictions. For $Z(t)$ we find that, after an initial Gaussian decay at very short times, the intermediate time decay is well described by a power law whose exponent is given precisely by the LM predictions. At longer times, beyond an $\alpha$-dependent time scale $t^*$, deviations occur. These deviations will be studied in detail in a future publication. For $e_{kin}(t)$ we find that the LM prediction works well even beyond the time scale $t^*$ associated with $Z(t)$. % What is important to note here is that the use of the kinetic Eq.~\eqref{eq:kinetic} allows to access longer times than known numerical methods.

\section{Evolution equation for higher order correlations}

To climb another step in the hierarchy of correlations we must switch to the Heisenberg representation:
\begin{equation}
  \partial_{t}O=i[H,O]\equiv iLO.
\end{equation}
Note the difference with the Liouvillian defined in~\eqref{eq:liouville}. The trace operation defines a dual projection operator over the observables of the Hilbert space:
\begin{equation}
\mathrm{Tr}\left[OP(t)\mu\right]=\mathrm{Tr}\left[\mu\mathsf{P}(t)O\right],
\end{equation}
where $O$  is an observable, $\mu$  a density matrix and $\mathsf{P}(t)$  is the observable space projection operator. In the Heisenberg representation the strategy is to separate the slow and fast components of the evolution operator $e^{iLt}$ using the projector $\mathsf{P}(t)$ and its complement $\mathsf{Q}(t)$. This allows to find an operator Langevin-like equation for the slow variables~\cite{grabert82_book_pot},
\begin{equation}
\partial_{t}\delta{n}(\mathbf{k},t)=\int_{0}^{t}ds\,\sum_{\mathbf{k}'}\Phi_{\mathbf{k},\mathbf{k}'}(s,t)\:\delta n(\mathbf{k}',s)+\eta_{\mathbf{k}}(t),
\end{equation}
where $\delta n(\mathbf{k},t)=e^{iLt}n(\mathbf{k})-\langle n(\mathbf{k})\rangle_{t}$. The memory function $\Phi_{\mathbf{k},\mathbf{k}'}(s,t)$ can be expressed in terms of the $\langle n(\mathbf{k})\rangle_{t}$'s. In particular, it is possible to show that~\cite{grabert82_book_pot}
\begin{equation}\label{eq:key}
\Phi_{\mathbf{k},\mathbf{k}'}(t,u)=\frac{\delta\int_{0}^{t}ds\, K_{\mathbf{k}}(t,s)}{\delta\langle n(\mathbf{k}')\rangle_{u}},
\end{equation}
where
\begin{equation}
K_{\mathbf{k}}(t,s)=-\alpha^{2}\mathrm{Tr}\left\{ [H_{1},e^{iH_{0}(t-s)}[H_{1},n(\mathbf{k})]e^{-iH_{0}(t-s)}]\sigma(s)\right\}+\mathcal{O}(\alpha^3).
\end{equation}
In the last equation a perturbative expansion analogous to that performed in the interaction representation has been used. The microscopic noise term $\eta_{\mathbf{k}}(t)$ cannot be written entirely in terms of the $\langle n(\mathbf{k})\rangle_{t}$'s. Again, we refer the interested reader to the literature to find an explicit expression for $\eta_{\mathbf{k}}(t)$~\cite{grabert82_book_pot} since for our immediate purposes we shall not need it. Using the Wick rule and taking the functional derivative Eq.~\eqref{eq:key} we find an explicit evolution equation for the time-correlation function of the slow variables $F_{\mathbf{k},\mathbf{k}'}(t)\equiv\mathrm{Tr}\left[\rho(0)\delta n(\mathbf{k},t)\delta n(\mathbf{k}',0)\right]$
\begin{equation}\label{eq:fluc}
F_{\mathbf{k},\mathbf{k}'}(t)=F_{\mathbf{k},\mathbf{k}'}(0)-16\alpha^{2}\int_{0}^{t}ds\,\left\{ F_{\mathbf{k},\mathbf{k}'}(s)A_{\mathbf{k}}(t,s)+\sum_{\mathbf{q}}F_{\mathbf{q},\mathbf{k}'}(s)\left[B_{\mathbf{k},\mathbf{q}}(t,s)-2C_{\mathbf{k},\mathbf{q}}(t,s)\right]\right\},
\end{equation}
where
\begin{eqnarray}
\nonumber B_{\mathbf{k},\mathbf{q}}(t,s)&=&\sum_{\mathbf{k}_{3},\mathbf{k}_{4}}\vert V^{\mathbf{k},\mathbf{q}}_{\mathbf{k}_{3},\mathbf{k}_{4}}\vert^{2}\frac{\sin\left[(t-s)\Delta e^{\mathbf{k},\mathbf{q}}_{\mathbf{k}_3,\mathbf{k}_4}\right]}{\Delta e^{\mathbf{k},\mathbf{q}}_{\mathbf{k}_3,\mathbf{k}_4}}\left(f(\mathbf{k},s)\bar{f}(\mathbf{k}_{3},s)\bar{f}(\mathbf{k}_{4},s)+\bar{f}(\mathbf{k},s)f(\mathbf{k}_{3},s)f(\mathbf{k}_{4},s)\right),\\
\nonumber C_{\mathbf{k},\mathbf{q}}(t,s)&=&\sum_{\mathbf{k}_{2},\mathbf{k}_{4}}\vert V^{\mathbf{k},\mathbf{k}_{2}}_{\mathbf{q},\mathbf{k}_{4}}\vert^{2}\frac{\sin\left[(t-s)\Delta e^{\mathbf{k},\mathbf{k}_2}_{\mathbf{q},\mathbf{k}_4}\right]}{\Delta e^{\mathbf{k},\mathbf{k}_2}_{\mathbf{q},\mathbf{k}_4}}\left(f(\mathbf{k},s)f(\mathbf{k}_{2},s)\bar{f}(\mathbf{k}_{4},s)+\bar{f}(\mathbf{k},s)\bar{f}(\mathbf{k}_{2},s)f(\mathbf{k}_{4},s)\right),
\end{eqnarray}
and $A_{\mathbf{k}}(t,s)=\sum_{\mathbf{k}'}B_{\mathbf{k}',\mathbf{k}}(t,s)$. To the best of our knowledge, it is the first time that Eq.~\eqref{eq:fluc} is derived. In order to solve Eq.~\eqref{eq:fluc} we need first to calculate the solution to Eq.~\eqref{eq:kinetic} to determine the coefficients $A_{\mathbf{k}}(t,s)$, $B_{\mathbf{k},\mathbf{q}}(t,s)$ and $C_{\mathbf{k},\mathbf{q}}(t,s)$. Once this is done, Eq.~\eqref{eq:fluc} is as amenable to numerical solution as Eq.~\eqref{eq:kinetic}. Solutions of this equation for specific 1D models shall be presented elsewhere. We remark here, however, that the Heisenberg representation POT could be useful to obtain evolution equations for other correlation functions using the same scheme.

\section{Conclusions}

We have presented a rigorous framework to obtain evolution equations for various observables and correlations functions in weakly interacting systems. These equations can be numerically solved in an efficient way. We have put to test the accuracy of the approach comparing with known results for 1D models. Moreover, the POT formalism could allow to obtain higher order corrections to these equations. The inherent limitation of the approach is that it is valid only in the perturbative regime.

\section*{Acknowledgments}

We acknowledge useful discussions with Miguel A. Cazalilla and the hospitality of the National Tsing Hua University in Taiwan, where part of the present work was realized. This work was supported by CONICET (PIP 0662), ANPCyT (PICT 2010-1907) and UNLP (PID X497), Argentina.

\end{document}